\documentclass[aps,prd,12point,twocolumn,nofootinbib,showpacs,superscriptaddress]{revtex4-1}

\usepackage{psfrag}
\usepackage{subfigure}
\usepackage{color}
\usepackage{mathrsfs}
\usepackage{graphicx}

\usepackage{amssymb}
\usepackage{amsmath, amsthm}
\usepackage{epstopdf}
\usepackage{hyperref}
\usepackage{enumerate}

\newcommand{\be}{\begin{equation}}
\newcommand{\ee}{\end{equation}}

\begin{document}

\title{Revisiting the Brans solutions of scalar-tensor
gravity}

\author{Valerio Faraoni}
\email[]{vfaraoni@ubishops.ca}
\affiliation{Physics Department and STAR Research Cluster,
Bishop's University, 2600 College St., Sherbrooke,
QC, Canada J1M~1Z7}

\author{Fay\c{c}al Hammad}
\email[]{fhammad@ubishops.ca}
\affiliation{Physics Department and STAR Research Cluster,
Bishop's University, 2600 College St., Sherbrooke,
QC, Canada J1M~1Z7}
\affiliation{Physics Department, Champlain
College-Lennoxville, 2580 College Street, Sherbrooke,
QC J1M 0C8, Canada}

\author{Shawn D. Belknap-Keet}
\email[]{sbelknapkeet02@ubishops.ca}
\affiliation{Physics Department,
Bishop's University, Sherbrooke, QC, Canada J1M~1Z7}


\begin{abstract}
Motivated by statements in the literature which contradict
two general theorems, the static and spherically symmetric
Brans solutions of  scalar-tensor gravity are analyzed
explicitly in both the Jordan and the Einstein conformal
frames. Depending on the parameter range, these solutions
describe wormholes or naked singularities but not
black holes.
\end{abstract}

\pacs{04.50.Kd, 04.20.Jb, 04.70.Bw}

\keywords{}

\maketitle

\section{Introduction}
\setcounter{equation}{0}

Brans-Dicke theory \cite{BD} is the prototypical
theory of gravity alternative to Einstein's General
Relativity (GR). Not long after its introduction, it was
generalized
to scalar-tensor theories \cite{ST} and, with the advent of
string theories, new interest was generated by the fact
that the simple bosonic string theory reduces to a
Brans-Dicke theory with coupling parameter $\omega=-1$
\cite{bosonic}. The original Brans-Dicke theory contains a
massless scalar field $\phi$ (acting approximately as the
inverse of the gravitational coupling strength
$\phi=G_\text{eff}^{-1}$)  and a dimensionless
parameter $\omega $ which would naturally be of order
unity,
but is constrained by Solar System experiments
to satisfy $| \omega| > 40000$ \cite{BertottiIessTortora}.
For this reason, theorists have moved on
to more sophisticated versions of Brans-Dicke theory, such
as scalar-tensor gravities \cite{ST} in which $\omega$
becomes a function of the Brans-Dicke scalar field, which
also acquires a mass or a self-interaction potential. In
cosmology,\footnote{Here ${\cal R}$ is the Ricci scalar
associated with the connection of
the spacetime metric $g_{ab}$.} $f({\cal R})$ theories of
gravity, which are ultimately classes of
scalar-tensor  theories with Brans-Dicke-like scalar
degree of freedom
$\phi=f'({\cal R})$, have
become extremely popular to explain the current
acceleration of the universe without invoking an {\em ad
hoc} dark energy (see the reviews \cite{SotiriouFaraoni,
DeFeliceTsujikawa, Odintsovreview}).  It is natural, in
this context,  to
search for
analogues of the Schwarzschild solution of GR. Shortly
after Brans-Dicke theory was introduced \cite{BD}, Brans
 presented
four families of geometries which are
static, spherically symmetric, vacuum solutions of the
Brans-Dicke field equations \cite{Brans}. Although there is
legitimate suspicion that these solutions may not be very
significant from the physical point of view (but the
literature has contradictory statements about this point),
it is
often necessary to pick some simple ({\em i.e.}, static,
spherical, and asymptotically flat) solutions of an
alternative theory of gravity as toy models for
theoretical purposes or  as
physical
solutions to test a theory experimentally. Currently a
large amount of
work is devoted to testing deviations from GR in black
hole environments (see, {\em e.g.}, \cite{BHtests}). The
Brans  solutions, being the first of their
kind discovered in scalar-tensor or dilaton gravity,
are a natural choice. However, they are surrounded
by some
ambiguity. According to a theorem by Agnese and La Camera
\cite{AgneseLaCamera}, all static and spherically symmetric
solutions of (Jordan frame) Brans-Dicke theory are either
naked singularities if the post-Newtonian parameter
\be
\gamma = \frac{\omega+1}{\omega+2}
\ee
satisfies $\gamma<1$, or wormholes if $\gamma  >1$. The
Brans classes~I-IV
solutions fall into this category and, therefore, they can
only describe naked singularieties or wormholes. This
result seems to be missed by several authors since
there are claims in the literature that certain static,
spherical classes of solutions of Brans-Dicke theory
describe black holes, which would contradict the Agnese-La
Camera theorem. For example, the Campanelli-Lousto
solutions \cite{CampanelliLousto} have been believed to
be black holes for a long
time until it was shown recently that
they indeed describe either wormholes or naked
singularities \cite{Vanzo}. Similarly, reading the
existing literature, because of explicit or implicit
statements one is left with the impression that
Brans
solutions can describe black holes for
some range of their parameters \cite{Bhattacharya09,
BhadraNandi2001, BhadraSarkar06, BhattaLaserra}. Similar
statements about ``cold
black holes'' similar to the Campanelli-Lousto solutions
are found in the literature \cite{BronnikovCold,
BhadraNandi2001, Yazadjiev03, BhadraSarkar06}. If Brans
geometries were black hole ones, they would also contradict
a theorem by Hawking \cite{Hawking} (recently extended to
general scalar-tensor gravity \cite{SotiriouFaraoniPRL,
Romano})
stating that all Brans-Dicke black holes are the same as in
GR.

Naked
singularities are of little
interest from the physical point of view because they
correspond to the breakdown of the Cauchy problem.
Wormholes are completely speculative objects
\cite{Visser}, but
there is plenty of astrophysical evidence for, and interest
in, black holes. It is of some interest, therefore, to
clarify the confusion existing in the literature about the
Brans geometries, which we set out to do.

The Brans-Dicke action in the absence of matter is
\be \label{BDaction}
S_\text{BD} = \int d^4x \, \frac{ \sqrt{-g}}{16 \pi}
\left(\phi  \mathcal{R}-\dfrac{\omega}{\phi}\nabla^a\phi
\nabla_a\phi \right)  \,,
\ee
where $\phi$ is the Brans-Dicke scalar field (approximately
equivalent to the inverse of the gravitational coupling),
${\cal R}$ is the Ricci scalar, and $g$ is the determinant
of the spacetime metric $g_{ab}$. We follow the notation
of Ref.~\cite{Waldbook}.  The Brans-Dicke field equations
{\em in vacuo} derived from the action~(\ref{BDaction}) are
\cite{BD}
\begin{eqnarray}
{\cal R}_{ab}-\frac{{\cal R}}{2} g_{ab} &=& \frac{\omega}{\phi^2}
\left( \nabla_a\phi \nabla_b\phi -\frac{1}{2}\, g_{ab}
\nabla^c \phi\nabla_c\phi \right)\nonumber\\
&&\nonumber\\
&\, & +\frac{1}{\phi}  \nabla_a\nabla_b\phi
\,,\label{BDfe}
\end{eqnarray}
\be
\Box \phi = 0 \,.
\ee
By performing the conformal transformation of the metric
\be \label{metric transformation}
g_{ab} \rightarrow \tilde{g}_{ab} = \phi \, g_{ab},
\ee
and the scalar field redefinition
\be
\phi \rightarrow \tilde{\phi}=\sqrt{\dfrac{|2\omega+3|}{16
\pi G }} \, \ln \left( \frac{\phi}{\phi_{*}}\right) \,,
\ee
where $\phi_{*}$ is a constant (Einstein frame quantities
are denoted by a tilde), the Brans-Dicke
action~(\ref{BDaction}) assumes its
Einstein frame form
\be
S_\text{BD} = \int d^4x\sqrt{-\tilde{g}}\left[
\dfrac{\tilde{\mathcal{R}}}{16 \pi}-\dfrac{1}{2}
\tilde{g}^{ab}\nabla_a\tilde{\phi} \nabla_b\tilde{\phi}
\right] \,.
\ee
This action formally looks like the Einstein-Hilbert action
of GR in the presence of a matter scalar field endowed with
canonical
kinetic energy. The Einstein frame vacuum field equations
are
\begin{eqnarray}
&&\tilde{\cal R}_{ab}-\frac{1}{2} \, \tilde{g}_{ab} \tilde{
{\cal
R}}= 8\pi G \left( \nabla_a \tilde{\phi} \nabla_b
\tilde{\phi} -\frac{1}{2} \, \tilde{g}_{ab}
\tilde{g}^{cd} \nabla_c \tilde{\phi}\nabla_d \tilde{\phi}
\right) \,,\nonumber\\
&& \label{Eframefe}\\
&& \tilde{g}^{ab} \tilde{ \nabla}_a  \tilde{ \nabla}_b
\tilde{\phi} =0 \,. \label{EframeKG}
\end{eqnarray}
We now proceed to analyze the four classes of Brans
solutions in the Jordan and in the Einstein conformal
frames.

\section{Brans class~I solutions}
\setcounter{equation}{0}

Class~I Brans solutions have been discussed in several
papers \cite{Nandi97, noted1, noted2, Nandi99,
BhadraSarkar06, BhadraNandic, Bhattacharya09,
Nandietal2008, VisserHochberg, LoboOliveira}.
It is found that these metrics
can describe wormholes, which is not
surprising since a Brans-Dicke-like scalar field in
scalar-tensor gravity has a non-canonical kinetic energy
and its effective stress-energy tensor on the right hand
side of eq.~(\ref{BDfe})  can
violate all of the energy conditions.
More recent solutions proposed in the
literature \cite{HeKim} have been identified as
special limits of Brans~I solutions \cite{BhadraSarkar06,
BhadraSimaciu}.

\subsection{Jordan frame}

In the Jordan frame representation, the Brans class~I line
element  and scalar
field are, respectively,
\begin{eqnarray}
ds^2_{(I)} &=& - \left( \frac{1-B/r}{1+B/r}
\right)^{2/\lambda} dt^2 \nonumber\\
&&\nonumber\\
&+ &     \left( 1+\frac{B}{r} \right)^4
\left( \frac{1-B/r}{1+B/r} \right)^{\frac{2(\lambda
-C-1)}{\lambda}} \left( dr^2 + r^2  d\Omega_{(2)}^2 \right)
\,,
\nonumber\\
&&\label{lineelementIJ}\\
\phi_{(I)} &=& \phi_0 \, \left( \frac{1-B/r}{1+B/r}
\right)^{C/\lambda}  \,,\label{BDscalar}
\end{eqnarray}
in polar coordinates $\left( t,r,\theta, \varphi
\right)$, where $r$ is an isotropic radius and
$d\Omega_{(2)}^2=d\theta^2 +\sin^2\theta \,
d\varphi^2$ is the line element on the unit 2-sphere. It
must be $r\geq 0$ and \cite{Brans}
\be\label{lambdasquared}
\lambda^2 =\left( C+1 \right)^2 -C\left( 1-\frac{\omega
C}{2} \right) >0 \,.
\ee
Here $\omega $ is a parameter of the theory and $B, C$, and
$\lambda$  are parameters of this family of solutions. $B$
plays the role of a mass parameter and, in analogy with
the
Schwarzschild geometry of GR, it makes sense to consider
only non-negative values of this parameter. There are actually two other reasons why
we restrict our study here to non-negative values of $B$. The first reason is that one can still
include the case $B=0$, but then class~I solutions simply reduce to the trivial Minkowski
space. The second reason is that one can also include the case $B<0$, but then, as we will shortly see, one just recovers
the case of positive $B$ by taking the mass parameter of the theory to be $-B$ instead of $B$. Therefore, we shall hereafter assume $B>0$.

Once the Brans-Dicke coupling parameter $\omega$ of the
theory is fixed, one has two independent parameters
$\left( B, C \right)$ or  $\left( B, \lambda \right)$,
since eq.~(\ref{lambdasquared}) relates $\lambda$ and $C$.
It is useful to restrict the parameter space and, to this
end, we note that, according to our assumption that $B>0$, only
positive values of $\lambda$ will be relevant here.
In fact, consider as an example the simple case
$C=0$, in which the Brans-Dicke scalar $\phi$ reduces to a
constant and eq.~(\ref{lambdasquared}) yields
$\lambda=\pm1$. When $\phi$ is constant, Brans-Dicke theory
reduces to GR and the Schwarzschild solution, which is the
unique vacuum, static, and spherically symmetric solution
of the Einstein equations must be recovered. By setting
$\lambda=1$ the line element~(\ref{lineelementIJ}) reduces
to
\be
ds^2_{(+1)} = -\left(
\frac{1-B/r}{1+B/r}
\right)^2dt^2 +
\left( 1+\frac{B}{r} \right)^4\!\left( dr^2 +r^2
d\Omega_{(2)}^2 \right) \label{linelCzero}
\ee
which is the Schwarzschild metric in isotropic
coordinates \cite{Bhattacharya09}. If instead $\lambda=-1$,
one obtains
\begin{eqnarray}
ds^2_{ (-1)} &=& -\left( \frac{1+B/r}{1-B/r}
\right)^2dt^2 \\
&&\nonumber\\
&\,& + \left( 1-\frac{B}{r} \right)^4 \left( dr^2 +r^2
d\Omega_{(2)}^2 \right).\label{negativeBMetric}
\end{eqnarray}
This is again just the Schwarzschild solution provided that, either $r\rightarrow-r$ which we don't consider here \cite{Bhattacharya09}, or that $B<0$ and, hence, as alluded to above, one then has just to interpret $-B$ as the mass parameter instead of $B$; a case we have already chosen to exclude. Therefore, we also assume $\lambda>0$ in the following.

We can now study the limit of Brans~I solutions to GR.
When $\omega \rightarrow \infty$, it is $\lambda^2 \simeq
\omega \, C^2/2 \rightarrow \infty $ and, if $C\neq 0$ (the
case $C=0$ having already been discussed), the
line element~(\ref{lineelementIJ}) reduces to
\be
ds_{(\infty)}^2 = -dt^2 +
\left( 1- \frac{B^2}{r^2} \right)^2
\left( dr^2 +r^2 d\Omega_{(2)}^2 \right) \,,
\ee
while the Brans-Dicke scalar becomes constant. If $C\neq
0$,  the corresponding solution of GR is not
recovered from Brans~I solutions in the $\omega \rightarrow
\infty $ limit. Instances in which
solutions of scalar-tensor
theories do not reduce to the corresponding GR
limit have been discussed in \cite{noGRlimit} and possible
reasons for
this behaviour have been identified in the
anomalous asymptotic dependence of $\phi$ on $\omega $
as $\omega \rightarrow \infty$ \cite{noGRlimit, reasons,
reasons2}.

The condition~(\ref{lambdasquared}) amounts to
imposing that $C$ is such that points on the parabola
of equation $\lambda^2(C)= \left( \frac{\omega}{2}+1
\right)C^2 +C+1$ lie in the $\lambda^2>0$ half-plane. The
roots of the equation $\lambda^2(C)=0$ are
\be
C_{\pm}= \frac{-1 \pm \sqrt{ -(2\omega+3)}}{\omega+2} \,.
\ee
and they are real only if $\omega \leq -3/2$. By looking at
the sign of the coefficient $ (1+\omega/2)$ of this
parabola, it is easy to establish that:

\begin{itemize}

\item If $\omega<-2$, the parabola has
concavity facing downward and intersects the $C$-axis at
$C_{\pm}>0$. It must be $C_{-}<C<C_{+}$.

\item If $\omega=-2$, then the parabola degenerates into
the straight line $\lambda^2=C+1$ and it must be $C>-1$.

\item If $-2< \omega < -3/2$, the parabola has concavity
facing upward and it must be $ C<C_{-}$ or $C>C_{+}$, where
$C_{-}<C_{+}<0$.

\item If $\omega=-3/2$, then
$\lambda^2=\left[(C+2)/2\right]^2 $ and the parabola has
concavity facing upward and touches the $C$-axis only at
$ C=-2$, therefore the only restriction is $C\neq -2$.

\item If $\omega >-3/2$, the concavity still faces upward
but there are no intersections between the
$C$-axis and the parabola, which always lies above it.
There is no restriction on the values of $C$.

\end{itemize}

Let us consider now the Ricci scalar ${\cal R}$: by
contracting the Brans-Dicke field equations~(\ref{BDfe})
and using
eq.~(\ref{BDscalar}), one obtains
\begin{eqnarray}
{\cal R} &=& \frac{\omega}{\phi^2} \, \nabla^c\phi
\nabla_c  \phi \nonumber\\
&&\nonumber\\
&=& \frac{4\omega
B^2C^2}{\lambda^2 r^4} \left[
\left( 1+\frac{B}{r} \right)^{-2-\frac{(C+1)}{\lambda} }
\left( 1-\frac{B}{r} \right)^{-2+\frac{(C+1)}{\lambda} }
\right]^2 \,.  \nonumber\\
&&\label{Ricci}
\end{eqnarray}
If $\omega\neq 0$ and $C\neq 0$, then the Ricci scalar is
singular at
$r=B$ when $ (C+1)/\lambda <2$. Whether this value of the
isotropic radius is physically significant is
discussed case-by-case below.

The areal radius is read off the line
element~(\ref{lineelementIJ}) and is
\be\label{arealradius}
R(r)=
\left( 1+\frac{B}{r} \right)^{ 1+\frac{(C+1)}{\lambda} }
\left( 1-\frac{B}{r} \right)^{ 1-\frac{(C+1)}{\lambda} }
r,
\ee
and its derivative is
\be\label{Rderivative}
\frac{dR}{dr}=  \left( \frac{ 1+B/r}{1-B/r} \right)^{
\frac{C+1}{\lambda} }
\left[ r^2 -\frac{2B(C+1)}{\lambda} \, r +B^2 \right] \,
\frac{1}{r^2} \,.
\ee
In the following it is useful to know the roots of the
equation $dR/dr=0$, which are
\be  \label{roots}
r_{(\pm)}= \frac{B(C+1)}{\lambda} \left( 1\pm \sqrt{
1-\left( \frac{\lambda}{C+1} \right)^2} \, \right) \,.
\ee
In order to make our discussion of the various
regions of the parameter space more compact, we focus on
the
possible values of the parameter combination
$(C+1)/\lambda$, which is relevant for both the roots of
the equation $dR/dr=0$ and in the search for horizons. The
horizons (which, when existing, are both apparent and
event horizons), are located by the roots of the equation
\cite{HernandezMisner, mylastbook}
\be
\nabla^cR \nabla_cR=0 \,,
\ee
which is equivalent to
\be
\left[ r^2 -2B\, \frac{(C+1)}{\lambda}r +B^2 \right]^2=0
\,.
\ee
Its roots coincide with those of the equation $dR/dr=0$
and, when they exist in the real domain, they are always
double roots. Let us
consider separately the various relevant cases.

\subsubsection{Parameter range $(C+1)/\lambda <1$}

In this case
\begin{eqnarray}
\frac{dR}{dr} &=&
\left( \frac{ 1+B/r }{
1-B/r} \right)^{\frac{C+1}{\lambda}  }
\left[ r^2 -2B\,
\frac{(C+1)}{\lambda} \, r +B^2 \right] \, \frac{1}{r^2}
\nonumber\\
&&\nonumber\\
&>& \left( 1+ \frac{B}{r} \right)^{ \frac{C+1}{\lambda}}
\left( 1-\frac{B}{r} \right)^{
-\frac{C+1}{\lambda} } \left( r-B \right)^2>0, \nonumber\\
&&
\end{eqnarray}
for all values of $r>B$. Moreover,
\be
R(r)= \left( 1+\frac{B}{r} \right)^{ 1+\frac{C+1}{r} }
\left( 1- \frac{B}{r} \right)^{ \left| 1-
\frac{C+1}{\lambda} \right|} r \,,
\ee
shows that $r=B$ corresponds to areal radius $R=0$,
hence the range $0< r<B $ is unphysical. The Ricci
scalar~(\ref{Ricci}) is singular at $R=0$. In this
parameter range the spacetime always hosts a naked central
singularity if $\omega \neq 0$. The details of the geometry
near this
singularity vary with the value of $(C+1)/\lambda $ as
described below.

\begin{itemize}

\item If $ 0< (C+1)/\lambda < 1$ then $dR/dr \rightarrow
+\infty$ as the spacetime singularity is
approached ($R\rightarrow 0^{+}$ (or  $r\rightarrow
B^{+}$).

\item If $ (C+1)/\lambda =0 $, then
\begin{eqnarray}
R(r) &=& \left( 1-\frac{B^2}{r^2} \right) r \rightarrow 0 \,,\\
&&\nonumber\\
\frac{dR}{dr} &=&  1+\frac{B^2}{r^2} \rightarrow  2
\,,
\end{eqnarray}
as the singularity at $R=0$ is approached.

\item If $ (C+1)/\lambda <0 $, then
\be
\frac{dR}{dr} =  \frac{ \left( 1-B/r\right)^{
\left| \frac{C+1}{\lambda}\right|} }{
\left( 1+B/r\right)^{  \left|
\frac{C+1}{\lambda} \right| } }
\left[ r^2 -2B\, \frac{C+1}{\lambda}\, r +B^2 \right]
\ee
tends to zero at the singularity $R=0$ or $r=B$.

\end{itemize}

\subsubsection{Parameter range $ (C+1)/\lambda =1 $}

We have $R(r)=r\left(1+B/r
\right)^2 $ and $R(B)=4B>0$, therefore the range $0<r<B$
of the isotropic radius is now physically meaningful.
Note that $R(r) \rightarrow +\infty $ as $r\rightarrow
0^{+}$ and that
\be
\frac{dR}{dr}= \left(1+\frac{B}{r} \right)
\left(1-\frac{B}{r} \right)  \,,
\ee
therefore the function $R(r)$ decreases if $0<r<B$,
has the absolute minimum $R(B)=4B>0$, and
increases for
$r>B$. The equation $\nabla^cR \nabla_cR=0$ locating the
horizons is equivalent to $ \left( 1-B/r \right)^2=0$,
with $r=B$ a double root. If $\omega \neq 0$, there is a
would-be wormhole throat at $r=B$ (or, at $R=4B$) where,
however, the Ricci scalar is  singular. This finite radius
singularity separates two disconnected spacetimes.

If $\omega=0$, $\lambda >0$, and $ 0<C<1$, also the
Brans-Dicke scalar diverges at $r=B$, which means that the
effective gravitational constant vanishes. If $C<0$ or
$C \geq 1$, then $\phi$ vanishes and the gravitational
coupling strength diverges. The case $C=0$ has already been
discussed for all values of $\omega$. In the context of
black holes, the divergence or the vanishing of the
Brans-Dicke scalar denotes ``maverick'' black holes which
are contrived, unstable, or pathological and are usually
discarded as unphysical ({\em e.g.},
\cite{SotiriouFaraoniPRL})
and the same criterion should be adopted for wormholes
(naked singularities are already unphysical).

\subsubsection{Parameter range $ (C+1)/\lambda >1 $}

In this case the equation
\begin{eqnarray}
\frac{dR}{dr} &=& \left( \frac{1+B/r}{1-B/r} \right)^{
\left| \frac{C+1}{\lambda} \right|}  \left[
r^2 -\frac{2B(C+1)}{\lambda} r +B^2 \right]\nonumber\\
&&\nonumber\\
&\,& \cdot \frac{1}{r^2} =  0
\end{eqnarray}
has the two roots~(\ref{roots}), which are both positive.
It is straightforward to see also that
\be
0< r_{(-)} <B< r_{(+)} \,.
\ee
The areal radius
\be
R(r) = \frac{
\left( 1+B/r \right)^{ \left|
1+\frac{C+1}{\lambda} \right|} r}{
\left( 1-B/r \right)^{\left|
\frac{C+1}{\lambda}-1
\right|}   } \rightarrow +\infty
\ee
as $r\rightarrow B^{+}$, hence the range $r<B$ of the
isotropic radius is unphysical and we ignore the root
$r_{(-)}<B$.
The apparent horizons are located at the roots of the
equation
\be
\left( 1-\frac{B}{r} \right)^{-2}
\left[ r^2 -\frac{2B(C+1)}{\lambda}r+B^2 \right]^2=0 \,.
\ee
Ignoring the root $r=B$, which corresponds to
$R=+\infty$, $r=r_{(+)}>B $ is a double root and we have
a wormhole throat at $r_{(+)}$. As seen earlier, the
Ricci scalar~(\ref{Ricci}) is singular at $r=B$ if
$(C+1)/\lambda <2$ and $\omega \neq 0$, but this
singularity is actually pushed to infinity since
$r\rightarrow B^{+}$ corresponds to infinite physical
radius $R$ hence this is an acceptable solution. The Ricci
scalar is regular for $(C+1)/\lambda \geq 2$.

\subsection{Einstein frame class~I solutions}

The Einstein frame metric and free scalar field are
\begin{eqnarray}
d\tilde{s}^2_{(I)} &= & \phi_{(I)} \, ds^2_{(I)}= -  \left(
\frac{1-B/r}{1+B/r}
\right)^{\frac{C+2}{\lambda} } dt^2 \nonumber\\
&&\nonumber\\
&+ &     \left( 1+\frac{B}{r}
\right)^{2+\frac{C+2}{\lambda} }
\left( 1-\frac{B}{r} \right)^{2-
\frac{(C+2)}{\lambda} } \left( dr^2 + r^2  d\Omega_{(2)}^2
\right)\nonumber\\
&& \label{lineelementIE}\\
\tilde{\phi}_{(I)} &=&  \sqrt{ \frac{|2\omega+3|}{16\pi G}
} \,\frac{C}{\lambda} \, \ln  \left( \frac{1-B/r}{1+B/r}
\right) +\mbox{const.} \label{BDscalarIE}
\end{eqnarray}
The areal radius and its derivative are
\begin{eqnarray}
\tilde{R}(r) &=& \phi \, R=
\left( 1+\frac{B}{r}\right)^{1+\frac{C+2}{2\lambda}}
\left( 1-\frac{B}{r}\right)^{1-\frac{C+2}{2\lambda}} r
\,,\nonumber\\
&&\\
\frac{d\tilde{R}}{dr} &=&
\left( \frac{1+B/r}{1-B/r} \right)^{\frac{C+2}{2\lambda}}
\left[ 1- \left( \frac{C+2}{\lambda} \right) \frac{B}{r}
+\frac{B^2}{r^2} \right] \,,\nonumber\\
&&
\end{eqnarray}
while the Einstein frame Ricci scalar (obtained by
contracting the field equations~(\ref{Eframefe})) is
\begin{eqnarray}
\tilde{ {\cal R}} &=& 8\pi G \, \tilde{g}^{rr} \left(
\frac{
d\tilde{\phi}}{dr} \right)^2 \nonumber\\
&&\nonumber\\
&=& \frac{2B^2C^2|2\omega+3|}{\lambda^2 r^4}
\left( 1+\frac{B}{r}\right)^{-4-\frac{C+2}{\lambda}}
\left( 1-\frac{B}{r}\right)^{\frac{C+2}{2\lambda}-4}
\,.\nonumber\\
&&
\end{eqnarray}
If $(C+2)/\lambda<4$, the Ricci scalar is
singular at $r=B$ (and it is always singular at $r=0$
unless $C=0$, in which case it is $\tilde{{\cal R}}=0$).

The equation $\nabla^c \tilde{R} \nabla_c
\tilde{R}=0$ locating the horizons is
\be
\left( 1- \frac{B^2}{r^2} \right)^{-2}  \left[ 1-
\left( \frac{C+2}{\lambda} \right) \frac{B}{r}
+\frac{B^2}{r^2} \right]^2=0
\ee
and has the same roots
\be
r_{(\pm)} =  \frac{B(C+2)}{2\lambda} \left( 1\pm \sqrt{
1-\frac{4\lambda^2}{(C+2)^2}} \,  \right)\label{EIroots}
\ee
as the equation $d\tilde{R}/dr=0$. When these roots exist
and are real and positive, they
are always double roots and, therefore, the solutions
always contain either wormhole throats or naked
singularities. Assuming that $B>0$ and $\lambda>0$ as in
the Jordan frame, if $ \left[ (C+2)/\lambda \right]^2<4$
there are no real roots and no horizons. If $
(C+2)/\lambda = \pm 2$  there is a quadruple root
$r_0=B$. If instead $ \left[ (C+2)/\lambda
\right]^2>4$, there are two double roots $r_{(\pm)}$.
Further, if $C>-2$ the two double roots $r_{(\pm)}$ are
both positive; if $C=-2$ the only (quadruple) root vanishes
and, if $C<-2$, there are no horizons. Let us examine the
situation in more detail.

\subsubsection{Parameter range $(C+2)/\lambda <-2$}

In this case it is
\be
\tilde{R}(r) =
\left( 1+\frac{B}{r} \right)^{1+\frac{C+2}{2\lambda} }
\left( 1-\frac{B}{r} \right)^{\left| 1+\frac{C+2}{2\lambda}
\right| } r
\ee
and $\tilde{R}(r) \rightarrow 0^{+}$ as $r\rightarrow B$,
hence the range $0<r<B$ is unphysical, while  $\tilde{R}(r)
\rightarrow 0^{+}$ as $r\rightarrow +\infty$. The roots
$r_{(\pm)}$ are negative and the Ricci scalar diverges at
$\tilde{R} =0 $, where there is a naked singularity.

\subsubsection{Parameter range $(C+2)/\lambda =-2$}

In this case
\be
\tilde{R}(r)=\frac{\left(r-B \right)^2}{r}
\ee
vanishes as $r\rightarrow B$ and diverges in both limits
$r\rightarrow 0^{+}$ and $r\rightarrow +\infty$.  It
could seem that there is a wormhole throat at $r=B$ but the
Ricci scalar diverges there. Also this geometry hosts a
naked singularity at $\tilde{R}=0$.

\subsubsection{Parameter range $-2< (C+2)/\lambda <2$}

Then
\be
\tilde{R}(r) =
\left( 1+\frac{B}{r} \right)^{1+\frac{C+2}{2\lambda} }
\left( 1-\frac{B}{r} \right)^{\left| 1-\frac{C+2}{2\lambda}
\right| } \frac{1}{r}
\ee
and
$\tilde{R}(r) \rightarrow 0^{+}$ as $r\rightarrow B$ while
$\tilde{R}(r) \rightarrow +\infty$ as $r\rightarrow
+\infty$. There are no real roots $r_{(\pm)}$ and no
horizons. The Ricci scalar diverges at $\tilde{R}=0$, where
we have again a naked singularity.

\subsubsection{Parameter range $(C+2)/\lambda =2$}

$r=B$ is is a quadruple root and
\be
\tilde{R}(r)= \left(1+\frac{B}{r} \right)^2 r
\ee
has the limits
$\tilde{R} \rightarrow +\infty$ as $r\rightarrow 0^{+}$ and
$\tilde{R} \rightarrow +\infty$ as $r\rightarrow +\infty$.
There is a wormhole throat at $\tilde{R}=4B$, the minimum
value of $\tilde{R}$.

\subsubsection{Parameter range $(C+2)/\lambda >2$}

Both double roots $r_{(\pm)}$ are positive and
\be
\tilde{R}(r)=
\left(1+\frac{B}{r} \right)^{1+\frac{C+2}{2\lambda} }
\left(1-\frac{B}{r} \right)^{-\left| 1-\frac{C+2}{2\lambda}
\right|} r
\ee
diverges in both limits $r\rightarrow B^{+}$ and
$r\rightarrow +\infty$, hence the range $0<r<B$ is
unphysical. In this parameter range it is
$0<r_{(-)}<B<r_{(+)}$ and there is a wormhole throat at
$r_{(+)}$.

\section{Brans class~II solutions}
\setcounter{equation}{0}

\subsection{Jordan frame class~II solutions}

There is a duality relating class II and class I solutions \cite{BhadraNandi2001,BhadraSarkar06}, so these two classes are not independent. We shall come back to this duality in Sect.~\ref{Dualities} below. The Jordan frame Brans class~II line element and scalar
field are
\begin{eqnarray}
ds^2_{(II)} &=& -\mbox{e}^{\frac{4}{\Lambda}\,
\arctan\left(
r/B \right)} dt^2 \nonumber\\
&&\nonumber\\
&+& \mbox{e}^{\frac{-4(C+1)}{\Lambda}\, \arctan\left(
r / B \right)} \left( 1+\frac{B^2}{r^2} \right)^2
\left( dr^2 +r^2 d\Omega_{(2)}^2 \right) \,,\nonumber\\
&&\label{IIlineelement}\\
\phi_{(II)} &=&\phi_0 \, \mbox{e}^{\frac{2C}{\Lambda}\,
\arctan\left( r/B \right)} \,,\label{IIBDscalar}
\end{eqnarray}
where
\be\label{IIconstraint}
\Lambda^2 = C\left( 1-\frac{\omega \,C}{2} \right) -\left(
C+1 \right)^2 >0.
\ee
This implies that $C\neq 0$, hence this value of the
parameter $C$ will not be considered in the following
even though it is clear that it would play a role if
the inequality~(\ref{IIconstraint}) is forgotten.
Indeed, note that if $\Lambda$ and $B$ are allowed to take simultaneously imaginary values, then setting $C=0$ will just turn the metric (\ref{IIlineelement}) into the Schwarzschild metric (\ref{linelCzero}) written in isotropic coordinates. We shall come back to this remark in Sect.~\ref{Dualities}.

Let us examine the possible range of the parameters $B,C$,
and $\Lambda$. The points of the parabola
$\Lambda^2(C)=-\left(  \frac{\omega}{2}+1 \right) C^2 -C-1$
must lie in the positive $\Lambda^2$ half-plane. This
parabola has concavity facing downwards and it intersects
the $C$-axis at
\be
C_{\pm}= \frac{ -1 \mp \sqrt{-\left(
2\omega+3\right)}}{\omega+2} \,.
\ee
There are no such intersections if $\omega>-3/2$ and two
coincident intersections if $\omega=-3/2$; we conclude that
it must be $ \omega <-3/2$ for class~II solutions to
exist in the Jordan frame. Assuming this condition, it is
easy to see that:

\begin{itemize}

\item If $-2< \omega <-3/2$ (corresponding to
$-(1+\omega/2)<0$), the parameter $C$ must lie in the range
$C_{-}<C<C_{(+)}$.

\item If $ \omega =-2$, the parabola degenerates into the
straight line $\Lambda^2(C) =-\left(C+1 \right)$ and it
must be $C <-1$.

\item If $ \omega <-2$, it must be
$C<C_{-}$ or $ C>C_{+}$.

\end{itemize}

The Ricci scalar is
\begin{eqnarray}
{\cal R} &=&  \frac{\omega}{\phi^2} \, \nabla^c \phi
\nabla_c  \phi \nonumber\\
&&\nonumber\\
&=& \frac{4\omega \, B^2 C^2 r^4\,
\mbox{e}^{\frac{4(C+1)}{\Lambda} \,
\arctan\left( r/B \right)}  }{\Lambda^2 \left(
r^2+B^2\right)^4} \nonumber\\
&&\nonumber\\
&=& \frac{4\omega \, B^2 C^2}{\Lambda^2} \,
\frac{   \mbox{e}^{\frac{4(C+1)}{\Lambda} \,
\arctan\left( r/B \right)} }{R^4}
\,.\label{IIRicci}
\end{eqnarray}
The only possible singularity of the Ricci scalar ${\cal
R}$ can occur as $R\rightarrow 0$. The areal radius is
\be \label{IIarealradius}
R(r) = \left( 1+\frac{B^2}{r^2} \right)
\mbox{e}^{\frac{-2(C+1)}{\Lambda}\, \arctan\left(
r/B \right)} r
\ee
and its derivative is
\be
\frac{dR}{dr}=\mbox{e}^{\frac{-2(C+1)}{\Lambda}\,
\arctan\left( r/B \right)} \left[ r^2-2B\,
\frac{(C+1)}{\Lambda}\, r+B^2 \right]  \,.
\ee
Note that $R>0 $ for all values of $r$ and
that $R\rightarrow +\infty$ as $r\rightarrow
+\infty$ and also as $r\rightarrow 0^{+}$.  Since the Ricci
scalar~(\ref{IIRicci}) can only diverge as $R\rightarrow
0^{+}$, there are no singularities of the Ricci scalar in
Brans class~II spacetimes.

The roots of the equation $dR/dr=0$ are
\be
r_{(\pm)} = \frac{B}{\Lambda} \left( C+1 \pm \sqrt{
(C+1)^2+ \Lambda^2} \, \right) \,.\label{IIroots}
\ee
Let us examine their sign, keeping in mind that
\begin{eqnarray}
\sqrt{ (C+1)^2+\Lambda^2} +C+1>0 \,,\\
\nonumber\\
C+1- \sqrt{ (C+1)^2+\Lambda^2} <0 \,.
\end{eqnarray}

\begin{itemize}

\item If $\Lambda B>0$, the parabola $\psi(r) \equiv
r^2-2B\frac{(C+1)}{\Lambda}\, r -B^2 $ has concavity facing
upwards and crosses the $r$-axis at $r_{(-)}$ and
$r_{(+)}$, with
$r_{(-)}<0< r_{(+)}$. Therefore $dR/dr<0$ and the function
$R(r)$ decreases if $0<r < r_{(+)}$, it has an absolute
minimum at $r_{(+)}$, and increases for $r>r_{(+)}$.

\item If $\Lambda B<0$, the parabola $\psi(r)$ still has
concavity facing upward but now $ r_{(+)}<0< r_{(-)}$ and
the discussion is the same as in the previous case provided
that the switch $ r_{(+)} \leftrightarrow r_{(-)}$ is made.

\end{itemize}

The equation $\nabla^c R \nabla_cR=0$ locating the horizons
becomes
\be
\left[ 1-\frac{B^2}{r^2} - \frac{2B(C+1)}{\Lambda} \, r
+B^2 \right]^2=0 \,.
\ee
The roots are the same as for the equation $dR/dr=0$
and,
when they are real and positive, they are always double
roots. This fact implies that there are no black hole
horizons and that class~II solutions do not
describe black holes but only wormhole throats or naked
singularities.  Further, the roots $r_{(\pm)}$ can be
written as
\be
r_{(\pm)} = \frac{B}{\Lambda} \left( C+1 \pm \sqrt{ C\left(
1-\frac{\omega \, C}{2}\right)} \, \right)
\ee
and the inequality~(\ref{IIconstraint}) implies that
$ C\left(1-\frac{\omega \,C}{2}\right) >\left( C+1
\right)^2 \geq 0$, hence there are always two real roots
$r_{(\pm)}$ of the equation $\nabla^cR \nabla_c R=0$
locating the horizons in the allowed range of parameters.
Are these roots positive? In order to answer this
question, note that $ \sqrt{ (C+1)^2 +\Lambda^2}+C+1>
\left| C+1 \right| +C+1 \geq 0$, hence
\be
\mbox{sign}\left( r_{(+)} \right) =\mbox{sign} \left(
\Lambda B \right) \,,
\ee
while $ C+1- \sqrt{ (C+1)^2 +\Lambda^2} <
C+1 - \left| C+1 \right|  \leq 0$ yields
\be
\mbox{sign}\left( r_{(-)} \right) =-\mbox{sign} \left(
\Lambda B \right) \,.
\ee
We can now analyze all the possibilities for the two
parameters $B$ and $\Lambda$.

\subsubsection{Parameter range $B>0$, ~$ \Lambda >0$}

When $B>0$ and $ \Lambda >0$, it is $ r_{(-)} <0<
r_{(+)} $and there is a double root $ r_{(+)}$ marking the
location of a wormhole throat. The same situation occurs
when $B<0$ and $ \Lambda <0$.

\subsubsection{Parameter range $B>0$, ~$ \Lambda <0$}

When $B>0$ and $ \Lambda <0$, it is $ r_{(+)} <0 <
r_{(-)} $ and there is a wormhole throat at $ r_{(-)}$. The
same situation occurs when $B<0$ and $ \Lambda >0$.

\subsubsection{Limit to GR}

Finally, let us consider the limit to GR  of
Brans~II
solutions. Since $\omega<-3/2$, the limit should be
$\omega\rightarrow -\infty$, which implies that
$ \Lambda^2 \approx -\omega \, C^2/2 \rightarrow +\infty $
(remember that $C\neq 0$). In this limit the Brans-Dicke
scalar~(\ref{IIBDscalar}) becomes constant but the
line element reduces to
\be
ds^2_{(\infty)}= -dt^2 +\left( 1+\frac{B^2}{r^2} \right)^2
\left( dr^2 +r^2 d\Omega_{(2)}^2 \right)  \,.
\ee
The areal radius is
\be
R(r) \approx r +\frac{B^2}{r} \,;
\ee
by inverting this relation one obtains $r^2-Rr+B^2=0$
and there
are the two values of the isotropic radius
\be
r_{1,2}=\frac{1}{2} \left( R\pm \sqrt{ R^2 -4B^2} \,
\right)
\ee
for each value of the physical areal radius $R$, which
implies that it must be $R\geq 2|B|$. The equation locating
the apparent horizons
\be
\nabla^c R\nabla_c R=g^{rr} \left(
\frac{dR}{dr} \right)^2= \left( \frac{
1-B^2/r^2}{1+B^2/r^2} \right)^2=0
\ee
has the double root $r=|B| $ (corresponding to $R=2|B|$).
There is always a wormhole throat in this spacetime, which
is not the spherically symmetric, static, asymptotically
flat, vacuum solution of GR ({\em i.e.}, Schwarzschild
space). Therefore, the limit $\omega
\rightarrow -\infty$ fails to reproduce the GR limit even
though the Brans-Dicke scalar becomes constant.

\subsection{Einstein frame class~II solutions}

The Einstein frame class~II line element and scalar field
are
\begin{eqnarray}
d\tilde{s}^2_{(II)} &=& \phi_{(II)} \, ds^2_{(II)}= -\mbox{e}^{\frac{6C}{\Lambda}\, \arctan\left( r/B
\right)} dt^2 \nonumber\\
&&\nonumber\\
&+& \mbox{e}^{\frac{-2(C+2)}{\Lambda}\, \arctan\left(
r/B \right)} \left( 1+\frac{B^2}{r^2} \right)^2
\left( dr^2 +r^2 d\Omega_{(2)}^2 \right) \,,\nonumber\\
&&\label{EFIIlineelement}\\
\tilde{\phi}_{(II)} &=& \sqrt{ \frac{|2\omega+3|}{16\pi G}
} \,
\frac{2C}{\Lambda} \,
\arctan\left( \frac{r}{B} \right)
+\mbox{const.} \label{EFIIscalar}
\end{eqnarray}
The areal radius and its derivative are
\begin{eqnarray}
\tilde{R}(r) &=& \left( 1+\frac{B^2}{r^2} \right)^2 \,
\mbox{e}^{\frac{-(C+2)}{\Lambda}\, \arctan\left(
r/B \right)} r\,,\\
&&\nonumber\\
\frac{ d\tilde{R}}{dr} &=&
\mbox{e}^{\frac{-(C+2)}{\Lambda}\, \arctan\left(
r/B \right)} \left[ 1-
\frac{(C+2)}{\Lambda}\,\frac{B}{r} -\frac{B^2}{r^2} \right]
\,,\nonumber\\
&&
\end{eqnarray}
while the Ricci scalar is
\be
\tilde{ {\cal R}} = \frac{2B^2 C^2 \left|2\omega +3
\right|}{\Lambda^2 r^4 \left( 1+B^2/r^2 \right)^4 } \,
\mbox{e}^{\frac{2(C+2)}{\Lambda}\, \arctan\left( r/B
\right)}
\ee
and is never singular. The equation $\nabla^c \tilde{R}
\nabla_c \tilde{R}=0$ locating the horizons is
\be
\left[ r^2- \frac{B(C+2)}{\Lambda}\,r
-B^2 \right]^2=0
\ee
and, when its roots
\be
r_{(\pm)}= \frac{B(C+2)}{\Lambda} \left[ 1\pm \sqrt{
1+\left( \frac{2\Lambda}{C+2}\right)^2} \, \right]\label{EIIroots}
\ee
are real and positive they are always double roots, hence
there can only be either wormhole throats or naked
singularities.

\subsubsection{Parameter range $B(C+2)/\Lambda <0$}

It is $r_{(+)}<0<r_{(-)}$ and
$\tilde{R}(r) \rightarrow +\infty $ as $r \rightarrow
0^{+}$, while
$\tilde{R}(r) \rightarrow +\infty $ as $r \rightarrow
+\infty$. There is a wormhole throat at $r_{(-)}$.

\subsubsection{Case $C=-2$}

It is $r_{(\pm)}=\pm B$ and $\tilde{R}(r)= \left( 1+B^2/r^2
\right)r$ has the limits
$\tilde{R}(r) \rightarrow +\infty $ as $r \rightarrow
0^{+} $ and
$\tilde{R}(r) \rightarrow +\infty $ as $r \rightarrow
+\infty$. There is a wormhole throat at $r=|B|$
(corresponding to physical radius $\tilde{R}=2|B|$).

\subsubsection{Parameter range $B(C+2)/\Lambda >0$}

In this case we have  $r_{(-)}<0<r_{(+)}$ and
$\tilde{R}(r) \rightarrow +\infty $ in both limits $r
\rightarrow 0^{+}$ and $r \rightarrow
+\infty$. There is a wormhole throat at $r_{(+)}$.

\section{Brans class~III solutions}
\setcounter{equation}{0}

Although it is claimed that
the class~III family does not admit wormholes \cite{Nandi97},
this is not the case, as shown below.

\subsection{Jordan frame class~III solutions}

The line element and Brans-Dicke scalar of Jordan frame
class~III Brans solutions are, respectively,
\begin{eqnarray}
ds^2_{(III)} &=& -
\mbox{e}^{-\frac{2r}{B}} dt^2
+\dfrac{B^4}{r^4} \, \mbox{e}^{\frac{2(C+1)}{B} \, r}
\left(
dr^2 +r^2 d\Omega_{(2)}^2 \right)
\,,\nonumber\\
&&\label{IIIlineelement}\\
\phi_{(III)} &=& \phi_0 \,\mbox{e}^{-Cr/B}
\,,\label{IIIBDscalar}
\end{eqnarray}
where
\be\label{IIIconstraint}
C=\frac{-1\pm \sqrt{ -\left( 2\omega+3\right)}}{\omega+2}
\ee
and, clearly $ B\neq 0, \omega \leq -3/2, \omega \neq -2$.
The areal radius and its derivative are
\begin{eqnarray}
R(r)&=& \frac{B^2}{r} \, \mbox{e}^{\frac{(C+1)}{B} \, r}
\,,\\
&&\nonumber\\
\frac{dR}{dr} &=& \frac{B\left( C+1 \right)}{r^2} \,
\mbox{e}^{ \frac{(C+1)}{B}\, r} \left( r-\frac{B}{C+1}
\right) \,.
\end{eqnarray}
We can rewrite the line element~(\ref{IIIlineelement})
using the areal radius $R$ instead of the isotropic radius
$r$ by means of the substitution
\be
dr=\frac{r^2}{B\left( C+1 \right) \left( r-\frac{B}{C+1}
\right)} \, \mbox{e}^{ -\frac{(C+1)}{B}\, r} dR \,,
\ee
which yields
\be
ds^2_{(III)} = - \mbox{e}^{-\frac{2r}{B}} dt^2
+\dfrac{B^2}{ \left(C+1\right)^2 \left(
r-\frac{B}{C+1}\right)^2} \, dR^2 +R^2 d\Omega_{(2)}^2  \,.
\ee
The horizons, when they exist, are located by the equation
$\nabla^cR\nabla_cR=0$, which becomes simply $g^{RR}=0$, or
\be
\left( \frac{C+1}{B} \right)^2 \left( r-\frac{B}{C+1}
\right)^2=0.
\ee
There is a double root
\be
r_\text{H}=\frac{B}{C+1}\label{JIIIrH}
\ee
when this quantity is positive, with corresponding areal
radius
\be
R_\text{H}=\mbox{e} B\left( C+1 \right) \,.
\ee
Therefore, there are either
zero or two coincident real roots and there can not be
black
holes: Brans class~III solutions always describe naked
singularities or wormholes.

The Ricci scalar is
\be
{\cal R}= \frac{\omega}{\phi^2} \, \nabla^c\phi
\nabla_c\phi = \frac{\omega \, C^2}{B^6} \,
\mbox{e}^{-\frac{2(C+1)}{B}\, r} \, r^4 \,.
\ee
Let us examine the various
possibilities for the range of parameters $B$ and $C$.

\subsubsection{Parameter range $C<-1$, ~$B>0$}

In this case we have
\begin{eqnarray}
R(r)&=& \dfrac{B^2}{r} \, \mbox{e}^{-\left|
\frac{C+1}{B} \right|  r} \,,\\
&&\nonumber\\
\frac{dR}{dr} &=& -\frac{B^2}{ r^2} \,
\mbox{e}^{-\left| \frac{C+1}{B}\right| r} \left(
1+\left| \frac{B}{C+1}\right|r\right)   \,,
\end{eqnarray}
and the function $R(r)$ is monotonically decreasing with
$R(r)\rightarrow 0$ as $r\rightarrow +\infty$ and
$ R(r)\rightarrow +\infty$ as $r\rightarrow 0^{+}$. Since
$R_\text{H}=\mbox{e} B\left(C+1\right)<0$ there are no
horizons and there
are no wormholes. The Brans-Dicke scalar $\phi=\phi_0
\mbox{e}^{ \left| \frac{C}{B} \right|r} \rightarrow 0 $ as
$R\rightarrow 0$ (corresponding to $r\rightarrow +\infty$)
and the Ricci scalar
\be
{\cal R} = \frac{\omega \,C^2}{B^6} \, \mbox{e}^{2\left|
\frac{C+1}{B} \right|r}
r^4 \rightarrow +\infty
\ee
as $R\rightarrow 0^{+}$. Therefore, there is a naked
singularity at $R=0$.

\subsubsection{Parameter range  $C>-1$, ~$B>0$}

In this case the areal radius $R(r) \rightarrow +\infty $
as $r\rightarrow +\infty$ and
$R(r) \rightarrow +\infty $
as $r\rightarrow 0^{+}$. Its derivative $dR/dr$ is
negative, and $R(r)$ decreases, for $0<r<r_\text{H}$.
$R(r)$ has
the absolute minimum $R_\text{H}=\mbox{e}B(C+1)>0$ at
$r_\text{H}$, and
increases for $r>r_\text{H}$. The double root $r_\text{H}$
of the
equation $\nabla^cR \nabla_cR=0$ is positive and there is a
wormhole throat at $r_\text{H}$, where the Brans-Dicke
field~(\ref{IIIBDscalar})
assumes the finite value $\phi_\text{H}=\phi_0 \,
\mbox{e}^{-\frac{C}{C+1} }$ and it becomes constant if
$C=0$ (the $C=0$ solution is treated below in the
discussion of the limit to GR).

Special subcases are:

\begin{itemize}

\item $C>0 $,~$B>0$, in which the Brans-Dicke scalar is a
finite and decreasing function of $r$ for all values of
this coordinate. Its derivative with respect to the areal
radius is
\begin{eqnarray}
\frac{d\phi}{dR}&=&
\frac{d\phi}{dr} \frac{dr}{dR}=
\frac{d\phi}{dr}\left( \frac{dR}{dr}\right)^{-1}
\nonumber\\
&&\nonumber\\
&=& \frac{ -\phi_0 C r^2}{B^2 (C+1)\left(r-r_\text{H}
\right)} \,
\mbox{e}^{-\frac{(2C+1)}{B} \, r} \rightarrow \infty
\end{eqnarray}
as $r\rightarrow r_\text{H}$. Therefore, for $r<r_\text{H}$
(or
$R>R_\text{H}$
in the ``left branch'' of $R$), it is $d\phi/dR>0$, with
$d\phi/dR \rightarrow +\infty$ as $r\rightarrow
r_\text{H}^{-}$.
For $r>r_\text{H}$ (or $R>R_\text{H}$ in its ``right
branch''),
instead,
it is $d\phi/dR<0$ with $d\phi/dR \rightarrow -\infty$ as
$r\rightarrow r_\text{H}^{+}$. The Brans-Dicke scalar has a
cusp,
but remains finite, at the horizon $R_\text{H}$ where it
attains
its maximum value, which means that the effective
gravitational coupling $G_\text{eff} \sim \phi^{-1}$ is
maximum
there).

\item $-1<C<0$,~~$B>0$: in this case the Brans-Dicke scalar
\be
\phi=\phi_0 \, \mbox{e}^{\left|\frac{C}{B}\right| r}
\ee
is an increasing function of the isotropic radius $r$ and
its derivative with respect to the areal radius is
\be
\frac{d\phi}{dR} =\frac{ |C|r^2}{B^2 \left(C+1\right)\left(
r-r_\text{H} \right)} \, \mbox{e}^{ \frac{-2(C+1)}{B}\, r}
\,.
\ee
We need to further distinguish the situation
$-1<C \leq -1/2$, in which
\be
\frac{d\phi}{dR} =\frac{ |C|r^2}{B^2 \left|
C+1\right|\left(
r-r_\text{H} \right)} \, \mbox{e}^{ \left|
\frac{2(C+1)}{B}\right|
r} \rightarrow \infty
\ee
as the wormhole throat is approached when $r\rightarrow
r_\text{H}$. In this case $d\phi/dR$ is negative for
$0<r<r_\text{H}$,
vanishes at $r_\text{H} $, and is positive for
$r>r_\text{H}$. The
Brans-Dicke scalar is minimum and finite, but has  a cusp
(and $G_\text{eff}$ is maximum) at the wormhole throat.

\end{itemize}

\subsubsection{Parameter range  $C>-1$, ~$B<0$}

In this case it is $R_\text{H}=\mbox{e}B\left( C+1\right)
<0$ and
there are no horizons and no wormhole throats. Since
$dR/dr$ is always negative the areal
radius is the decreasing function of the isotropic radius
\be
R(r)=\frac{B^2}{r} \, \mbox{e}^{-\left| \frac{C+1}{B}
\right|r} \,.
\ee
The limit $r\rightarrow 0^{+}$ corresponds to $R\rightarrow
+\infty$, while $r\rightarrow +\infty$
corresponds to $R\rightarrow 0^{+}$. The Ricci scalar is
\be
{\cal R}= \frac{\omega \, C^2}{B^6} \,\mbox{e}^{
2\left| \frac{C+1}{B}\right| r} \rightarrow +\infty
\ee
as $R\rightarrow 0^{+}$. Therefore, there is central naked
singularity for these parameter values.

\subsubsection{Parameter range  $C=-1$, ~$B\neq0$}

In this case we have $R=B^2/r$ and the line element becomes
\begin{eqnarray}
ds^2_{(III)} &=& -\mbox{e}^{ -2r/B } dt^2 +\frac{B^4}{r^4}
\left( dr^2+r^2d\Omega_{(2)}^2 \right) \nonumber\\
&&\nonumber\\
&=&  -\mbox{e}^{-2r/B } dt^2
+dR^2+R^2d\Omega_{(2)}^2 \,.
\end{eqnarray}
The spatial sections are flat and there are no horizons.
The limits $r\rightarrow 0^{+}$ and $r\rightarrow +\infty $
corresponds to $R\rightarrow +\infty$ and  $R\rightarrow
0^{+}$, respectively. Both the Ricci scalar and the
Brans-Dicke  scalar field
\begin{eqnarray}
{\cal R} &=& \frac{\omega \,B^2}{R^4} \,,\\
&&\nonumber\\
\phi &=& \phi_0 \, \mbox{e}^{B/R} \,,
\end{eqnarray}
diverge as $R\rightarrow 0^{+}$: there is a naked
singularity at $R=0$.

\subsubsection{Parameter range  $C<-1$, ~$B<0$}

This situation is identical to the case  $C>-1$, ~$B>0$.

\subsubsection{Limit to GR}

Finally, let us discuss the limit to GR $\omega\rightarrow
-\infty$, which yields $C\rightarrow 0$. In this limit the
Brans-Dicke scalar~(\ref{IIIBDscalar}) becomes constant and
the line element reduces to
\be
ds^2_{(\infty)}= -\mbox{e}^{-2r/B} dt^2
+\left( \frac{B}{r-B} \right)^2 dR^2 +R^2 d\Omega_{(2)}^2
\,.
\ee
There is a wormhole throat at the horizon
$R=R_\text{H}=\mbox{e}B$. Also for Brans~III solutions, the
limit
in which $\phi$ becomes constant does not reproduce the
corresponding solution of GR.

\subsection{Einstein frame class~III solutions}

In the Einstein frame the line element and
scalar of class~III solutions are, respectively,
\begin{eqnarray}
d\tilde{s}^2_{(III)} &=& \phi_{(III)} \, ds^2_{(III)}
\nonumber\\
&&\nonumber\\
&=&  -
\mbox{e}^{-\frac{(C+2)r}{B}} dt^2
+\dfrac{B^4}{r^4} \, \mbox{e}^{\frac{(C+2)r}{B}}
\left( dr^2 +r^2 d\Omega_{(2)}^2 \right)
\,,\nonumber\\
&&\label{EFIIIlineelement}\\
\tilde{\phi}_{(III)} &=& - \sqrt{ \frac{
|2\omega+3|}{16\pi G} }
\,\frac{Cr}{B} +\mbox{const.}   \label{EFIIIscalar}
\end{eqnarray}
The areal radius and its derivative are
\begin{eqnarray}
\tilde{R}(r) &=& \frac{B^2}{r} \,
\mbox{e}^{\frac{(C+2)r}{2B}} \,,\\
&&\nonumber\\
\frac{d\tilde{R}}{dr} &=& \frac{B^2}{r^2} \,
\mbox{e}^{\frac{(C+2)r}{2B}} \left( \frac{C+2}{2B} \right)
\left( r-\frac{2B}{C+2} \right), \nonumber\\
&&
\end{eqnarray}
while the Einstein frame Ricci scalar is
\be
\tilde{ {\cal R}}=\frac{ |2\omega+3|C^2r^4}{2B^6} \,
\mbox{e}^{-\frac{(C+2)r}{B}}  \,.
\ee
The equation $\nabla^c \tilde{R} \nabla_c \tilde{R} =0$
becomes
\be
\left(C+2\right)^2 \left( r-\frac{2B}{C+2} \right)^2=0
\ee
and has the double root
\be
r_* =\frac{2B}{C+2}.\label{EIIIrH}
\ee
(if $C=-2$ there are no roots).

\subsubsection{Parameter range $B/(C+2)>0$}

In this range of parameters the double root $r_*$ is
positive and the areal radius $\tilde{R}(r)$ diverges in
both limits $r\rightarrow 0^{+}$ and $r\rightarrow
+\infty$. There is a wormhole throat at $r_*$,
corresponding to $\tilde{R}_*=\mbox{e} B(C+2)/2$.

\subsubsection{Parameter range $B/(C+2)<0$}

In this case there are no horizons, the areal radius
$\tilde{R}(r)$ tends to zero value
as $r\rightarrow
+\infty$, where the Ricci scalar diverges,
 and to infinity as $r\rightarrow 0^{+}$. There is a naked
central singularity.

\subsubsection{Case  $C =-2$}

In this case there are no horizons and the areal radius
$\tilde{R}(r)=B^2/r $ behaves as in the previous case. The
Ricci scalar diverges again at $\tilde{R}=0$ and there is a
naked central singularity.

\section{Brans class~IV solutions}
\setcounter{equation}{0}

There is another duality relating class~III
and class~IV solutions \cite{BhadraSarkar06}. We shall come back again to this duality in Sect.~\ref{Dualities}. Brans~IV
solutions were examined, for a restricted range of
parameters,\footnote{Specifically, for the situations $B>0$
; and  $ B>0, C>-1$.} in the
recent Ref.~\cite{FaraoniPrainMoreno} in both the Jordan
and Einstein frames. There it is shown that, for a certain
range of parameters,  the formal solution is a wormhole
in the Jordan frame and a naked singularity in the
Einstein frame, and the detailed reason why this happens
was pointed out \cite{FaraoniPrainMoreno}. For
completeness, we briefly revisit also those cases.

\subsection{Jordan frame class~IV solutions}

The Jordan frame line element and Brans-Dicke scalar field
for Brans class~IV solutions are, respectively\footnote{Note that in the original Brans' class~IV metric and the corresponding scalar field, introduced in Ref.~\cite{Brans}, the parameter $B$ appears in the denominator of the exponents and, hence, has there the dimensions of inverse length. We chose here to put $B$ in the numerators in order for it to have the same dimensions of length as it does within the other three classes. Furthermore, it is only under these forms of the metric and the scalar field that the dualities we are going to discuss in Sect.~\ref{Dualities} appear to be more than just mathematical transformations of the label $r$.},
\begin{eqnarray}
ds^2_{(IV)} &=& - \mbox{e}^{-\frac{2B}{r}} dt^2 +\mbox{e}^{
\frac{2B(C+1)}{r}} \left( dr^2 +r^2 d\Omega_{(2)}^2 \right)
\label{IVlineelement}\\
&&\nonumber\\
\phi_{(IV)} &=& \phi_0 \, \mbox{e}^{ -\frac{BC}{r}} \,,
\label{IVBDscalar}
\end{eqnarray}
where
\be\label{IVconstraint}
C=\frac{ -1\pm \sqrt{ -\left( 2\omega+3\right)}}{\omega+2}
\,.
\ee
Clearly, the Brans-Dicke parameter is limited to the range
$\omega \neq -2, \omega<-3/2$. The parameter $B$ has the
dimensions of a length and $r>0$. The areal radius
and its derivative are
\begin{eqnarray}
R(r) &=&\mbox{e}^{\frac{B(C+1)}{r} } r  \,,\\
&&\nonumber\\
\frac{dR}{dr}&=& \mbox{e}^{\frac{r_\text{H}}{r} }\left(
1-\frac{r_\text{H}}{r} \right) \,,
\end{eqnarray}
where
\be
r_\text{H}= B(C+1)  \label{IVrH}
\ee
is the root of the equation $dR/dr=0$ and
$R_\text{H}=\mbox{e}\,r_\text{H} $
is the corresponding value of the areal radius. The
isotropic radius (\ref{IVrH}) is also the double root of
the equation locating the horizons $\nabla^cR\nabla_cR=0$,
which becomes $\left( 1-r_\text{H}/r \right)^2=0$. When
$r_\text{H}$ is
real and positive the solution describes a wormhole,
otherwise there are no horizons and no black holes. The
Ricci scalar is
\be
{\cal R}= \frac{ \omega}{\phi^2} \, \nabla^c\phi
\nabla_c\phi = \frac{\omega \, B^2C^2}{r^4} \,
\mbox{e}^{-\frac{2B(C+1)}{r} } \,.
\ee

\subsubsection{Parameter range $B>0, C>-1$}

In this case $r_\text{H}=\left| B(C+1) \right|>0$ and the
areal
radius $R(r) = \mbox{e}^{ \left| \frac{B(C+1)}{r} \right|}r$
diverges as $r\rightarrow 0^{+}$ and as $r\rightarrow
+\infty$; it decreases for $0<r<r_\text{H} $, assumes its
minimum
value at $r_\text{H}$ and increases for $r>r_\text{H}$. We
have a
wormhole throat at $R_\text{H}=\mbox{e} B(C+1)$, at
which $\phi$ and ${\cal R}$
are finite (see \cite{FaraoniPrainMoreno} for further
discussion).

\subsubsection{Parameter range $B<0, C>-1$}

The root $r_\text{H}=-\left| B(C+1) \right|$ is negative
and
there are no horizons. The areal radius $R(r)$
increases monotonically from  zero value as $r\rightarrow
0^{+}$
reaching infinity as $r\rightarrow +\infty$. The Ricci
scalar
\be
{\cal R}= \frac{\omega \, B^2C^2}{r^4} \, \mbox{e}^{2\left|
\frac{B(C+1)}{r} \right|}
\ee
diverges as $r\rightarrow 0{+}$
(or $R\rightarrow 0^{+}$), signaling a central naked
singularity.

\subsubsection{Parameter range $B<0$, ~$C<-1$}

The double root $r_\text{H}=\left|B(C+1)\right|$ is
positive,
$R(r)\rightarrow +\infty$ as $r\rightarrow 0^{+}$ and as
$r\rightarrow +\infty$. There is a wormhole throat at
$r_\text{H}$, with $\phi$ and ${\cal R}$ finite.

\subsubsection{Parameter range $ B>0$, ~$C<-1$}

It is $r_\text{H}=-\left|B(C+1)\right|<0$ and there are no
horizons. The areal radius $R(r)$ increases monotonically
from zero value at $r=0$ to infinity as $r\rightarrow
+\infty$. The Ricci scalar
\be
{\cal R}= \frac{\omega \, B^2C^2}{r^4} \,
\mbox{e}^{ 2\left| \frac{B(C+1)}{r} \right|}
\ee
diverges as $r\rightarrow 0^{+}$ ({\em i.e.}, as
$R\rightarrow 0^{+}$), signaling again a central naked
singularity.

\subsubsection{Parameter range $C=-1$, ~$B \neq 0$}

This situation corresponds to $\omega=-2$, which is
excluded by eq.~(\ref{IVconstraint}). However, one could
think of considering the formal line
element~(\ref{IVlineelement}) without reference to its
derivation in \cite{Brans}, that is,
\be
ds^2_{(IV)}=- \mbox{e}^{  -\frac{2B}{r} }dt^2 + dr^2+r^2
d\Omega_{(2)}^2,
\ee
for which areal and isotropic radius coincide, and which
has flat spatial sections. The Ricci scalar
\be
{\cal R}= \frac{\omega \, B^2C^2}{R^4}
\ee
diverges as $R\rightarrow 0$ and there is a naked
singularity.

\subsubsection{The GR limit}

The GR limit should correspond to $\omega \rightarrow
-\infty$, which implies that $C\rightarrow 0$ and
$r_\text{H}
\rightarrow B$. The scalar field becomes constant in this
limit, the line element reduces to
\be ds^2_{(\infty)}= -\mbox{e}^{\frac{-2B}{r}} dt^2 +
\mbox{e}^{\frac{2B}{r}}\left( dr^2 +r^2 d\Omega_{(2)}^2
\right) \,,
\ee
and the areal radius is $R(r)=r\mbox{e}^{\frac{B}{r}}$.
The Ricci scalar is
\be
{\cal R}= \frac{\omega \, B^2C^2}{r^4} \,
\mbox{e}^{-\frac{2B(C+1)}{r} } \approx
\frac{-2B^2}{r^4} \,
\mbox{e}^{-\frac{2B}{r} } \,.
\ee
If $B>0$, it is $r_\text{H}>0$, $R(r) \rightarrow +\infty$
as
$r\rightarrow 0^{+}$ and as $r\rightarrow +\infty$. The
Ricci scalar is finite at $R=0$ and the solution describes
a wormhole.

If $B<0$, it is $r_\text{H}<0$ and there are no horizons.
the
areal radius $R(r) \rightarrow 0 $ as
$r\rightarrow 0^{+}$ and
$R(r) \rightarrow+\infty$ as $r\rightarrow +\infty$. The
Ricci scalar diverges at $R=0$: there is a central naked
singularity.

In either case the limit in which $\phi$ becomes constant
fails to reproduce the corresponding GR solution.

\subsection{Einstein frame class~IV solutions}
\label{subsec:}

The Einstein frame line element and  scalar
field for class~IV solutions are
\begin{eqnarray}
d\tilde{s}^2_{(IV)} &=& \phi_{(IV)} \, ds^2_{(IV)}
\nonumber\\
&&\nonumber\\
\,& =& - \mbox{e}^{-\frac{B(C+2)}{r}}
dt^2 +\mbox{e}^{ \frac{B(C+2)}{r}}
\left( dr^2 +r^2 d\Omega_{(2)}^2 \right)
\,,\nonumber\\
&&\label{EFIVlineelement}\\
\tilde{\phi}_{(IV)} &=& \sqrt{ \frac{ |2\omega+3|}{16\pi G}
} \,
\frac{BC}{r}  +\mbox{const.}
\label{EFIVscalar}
\end{eqnarray}
The areal radius is simply
\be
\tilde{R}=\mbox{e}^{\frac{B(C+2)}{2r}}
\ee
and the Ricci scalar is
\be
\tilde{{\cal R}} = \frac{ |2\omega +3 |B^2C^2}{2r^4} \,
\mbox{e}^{\frac{-B(C+2)}{r}} \,.
\ee
The equation $\nabla^c \tilde{R} \nabla_c \tilde{R}=0$
becomes  $\left( r-r_* \right)^2=0$, where
\be
r_*= \frac{B(C+2)}{2}\label{EIVrH}
\ee
is the only root and a double root.

\subsubsection{Parameter range $B(C+2)<0$}

In this case $r_*<0$ and there are no horizons. The areal
radius has the limits $\tilde{R}(r) \rightarrow 0^{+}$ as
$r \rightarrow 0^{+}$, where the Ricci scalar diverges, and
$\tilde{R}(r) \rightarrow +\infty$ as $r \rightarrow
+\infty$. There is a naked central singularity.

\subsubsection{Parameter range $B(C+2)>0$}

In this case $r_*>0$, and $\tilde{R}(r) \rightarrow
+\infty$ in both limits $r \rightarrow 0^{+}$ and
$r\rightarrow +\infty$. There is a wormhole throat at
physical radius $\tilde{R}_{*}= \mbox{e}B(C+2)/2$.

\subsubsection{Case $C=-2$}

In this case areal radius and isotropic radius coincide and
the Ricci scalar diverges as $\tilde{R} \rightarrow 0^{+}$.
There is
a naked central singularity.

\section{The dualities}\label{Dualities}
As mentioned above, Brans solutions are not actually all independent as there are dualities relating pairs of the solution classes \cite{BhadraNandi2001, BhadraSarkar06}. There is a duality relating classes I and II and there is another duality relating classes III and VI. It is therefore not surprising to find the same pattern concerning the existence of wormholes and/or naked singularities within a pair of solutions related by such a duality. It is also not a coincidence that all the Brans classes fail to recover the GR limit as $\omega\rightarrow\infty$.

Furthermore, as we shall see shortly, these dualities are akin to the duality one finds for the Schwarzschild black hole solution when the latter is written in isotropic coordinates. Indeed, it is well known that the Schwarzschild metric in isotropic coordinates is self-dual under the inversion $r\leftrightarrow B^2/r$, as it can easily be verified using the metric (\ref{linelCzero}). This fact might actually have been expected as the Schwarzschild solution in isotropic coordinates is recovered either from class I or class II when the parameter $C$ vanishes as we saw in eq.~(\ref{linelCzero}) and in the remark below eq.~(\ref{IIconstraint}), respectively, for then the Brans-Dicke field $\phi$ becomes a constant. The same observation also applies to the case of classes III and IV, as the latter reduces to the Minkowski spacetime for $B=0$ while the former reduces, for $B\rightarrow\infty$ ($C$ cannot vanish for these classes), to a Minkowski spacetime whose radial coordinate $r$ has been inverted to $1/r$. All these observations remain valid when going to the Einstein frame.

The duality transformation that relates class III to class IV is the following inversion,
\begin{equation}\label{DualityIII-VI}
r\longleftrightarrow\frac{B^{2}}{r}.
\end{equation}
Notice that the form of the duality transformation displayed here is slightly different from that given in Ref.~\cite{BhadraSarkar06}. The dimensions here are correct as the parameter $B$ has the same dimensions of length as $r$. In fact, by a straightforward substitution, one easily recovers in the Jordan frame the metric (\ref{IIIlineelement}) from the metric (\ref{IVlineelement}), and {\em vice-versa}, thanks to this transformation of the coordinate $r$. In the Einstein frame one also recovers the metric (\ref{EFIVlineelement}) from the metric (\ref{EFIIIlineelement}) using such an inversion. The effect of this inversion is easily seen by comparing the location of the would-be wormholes' throats (\ref{JIIIrH}) and (\ref{IVrH}) in the Jordan frame and (\ref{EIVrH}) and (\ref{EIIIrH}) in the Einstein frame; one just being the inverse of the other up to the factor $B^{2}$ as dictated by the inversion (\ref{DualityIII-VI}).

The duality transformation that relates class I to class II is,
\begin{equation}\label{DualityI-II}
r\longleftrightarrow\frac{B^{2}}{r},\qquad\lambda\longleftrightarrow-i\Lambda,\qquad B\longleftrightarrow iB.
\end{equation}
Here also, our notation differs from that of Refs.\cite{BhadraNandi2001,BhadraSarkar06} in that we used $B^2$ for the $r$-inversion in order to get the dimensions right. In fact, by a straightforward substitution, one easily recovers in the Jordan frame the metric (\ref{IIlineelement}) from the metric (\ref{lineelementIJ}), and {\em vice-versa}, thanks to these three transformations. The same applies to the metrics (\ref{lineelementIE}) and (\ref{EFIIlineelement}) in the Einstein frame.

In contrast to the case of classes III and IV, however, the effect of the duality transformation (\ref{DualityI-II}) is not to make the radii (\ref{roots}) and (\ref{IIroots}) of the would-be wormhole throats in the Jordan frame inverse of each other. The same applies to the case of the radii (\ref{EIroots}) and (\ref{EIIroots}) of the would-be wormholes in the Einstein frame. In both cases, one is recovered from the other just by making the substitutions (\ref{DualityI-II}) on $\Lambda$ and $B$ but the resulting radii are not inverse of each other. This could easily be understood by the fact that, in contrast to classes III and IV which admit as a limit the Minkowski spacetime, classes I and II admit as a limit the Schwarzschild metric which is already self-dual under the inversion $r\leftrightarrow B^2/r$ in isotropic coordinates.

Now, since the parameter $B$ has the dimensions of length and the parameters $\lambda$ and $\Lambda$ both appear in exponents inside the metrics (\ref{IIlineelement}) and (\ref{lineelementIJ}), it might seem unphysical to change these parameters into imaginary entities. Note, however, that the meaning we give to the parameter $B$ cannot be taken independently from the parameter $\lambda$ in class I or independently from the parameter $\Lambda$ in class II. In fact, as we saw in eq.~(\ref{linelCzero}), for $C=0$ the parameter $B$ becomes the mass parameter of the Schwarzschild solution if $\lambda=1$, whereas for $\lambda=-1$ it is the parameter $-B$ that should be interpreted as the Schwarzschild mass parameter in eq.~(\ref{negativeBMetric}). The same applies for class II whose metric (\ref{IIlineelement}) reproduces the Schwarzschild metric in isotropic coordinates only when both $\Lambda$ and $B$ become imaginary.

Finally, one might wonder at this point if there still exists another duality transformation that might relate one pair of classes to another pair. The answer is no and the reason is the following. The fact that one pair of solutions (classes I and II) reduces to the Schwarzschild metric for a specific value of the parameter $C$ and the other pair (classes III and IV) reduces to the Minkowski spacetime for a specific value of the parameter $B$ is what forbids the existence of any duality between the two pairs. In other words, as there is no duality between the curved Schwarzschild spacetime and the flat Minkowski spacetime, no duality could exist between the first pair and the second pair either.

\section{Conclusions}
\setcounter{equation}{0}

We have verified explicitly that the Brans classes~I-IV of
solutions
\cite{Brans} of Brans-Dicke theory \cite{BD} always
describe either wormholes or else horizonless geometries
containing
naked singularities, and they never describe black holes,
in agreement with the Agnese-La~Camera theorem
\cite{AgneseLaCamera} and with Hawking's theorem on
Brans-Dicke black holes \cite{Hawking,
SotiriouFaraoniPRL, Romano}.
Ambiguity and confusion lingering
in the literature about the nature of these solutions and
apparent contradictions with the theorems mentioned
above are thus eliminated
once and for all. Spacetimes harboring naked singularities
are unphysical since one cannot prescribe regular initial
data in the presence of a naked singularity and the initial
value problem fails, leaving the theory void of
predictability and wormholes as the only
remaining Brans
solutions. Wormholes are exotic objects which require the
energy conditions to be violated. The Brans solutions are
vacuum solutions and the Brans-Dicke scalar acts as the
only form of effective ``matter'' once the field equations
are written as
effective Einstein equations, as in eq.~(\ref{BDfe}). It is
well known that a
non-minimally coupled scalar field and the Brans-Dicke
scalar can violate all of the energy conditions, therefore
it is no surprise that one can obtain wormholes as
solutions of vacuum Brans-Dicke theory, as has already been
remarked in the literature.

No Jordan frame solution of any Brans class has the correct
limit to GR, which lends a word of caution on using these
solutions as toy models, since they are rather
pathological. The reason for their failure to
reproduce the corresponding GR solution ({\em i.e.}, the
Schwarzschild geometry) is by now well understood, as
recalled above.

The Einstein frame versions of the Brans solutions,
naturally, describe only spacetime geometries containing
either wormholes or naked singularities. It is interesting
that a wormhole can sometimes be conformally transformed to
a naked singularity or {\em vice-versa}. These instances
and the general reasons why they occur were analyzed in
detail in the recent Ref.~\cite{FaraoniPrainMoreno}, using
Brans~IV
solutions as an example. The present work provides more
examples.

\begin{acknowledgments}
This work is supported by the Natural Sciences and
Engineering Research Council of Canada.
\end{acknowledgments}

\end{document}